\documentclass[a4paper,11pt]{article}
\usepackage{bm}
\usepackage{amsmath}

\begin{document}

\begin{center}

\large
{\bf Microscopic Calculation of Spin Torques and Forces}

\vskip 5mm
\normalsize
{H. Kohno$^a$ \footnote{kohno@mp.es.osaka-u.ac.jp}, 
G. Tatara$^{b,c}$, J. Shibata$^d$, Y. Suzuki$^a$}

\vskip 3mm
\small

$^a$ {\it Graduate School of Engineering Science, 
Osaka University, Toyonaka, Osaka 560-8531, Japan} \\
$^b$ {\it Graduate School of Science, Tokyo Metropolitan University, 
Hachioji, Tokyo 192-0397, Japan} \\
$^c$ {\it PRESTO, JST, 4-1-8 Honcho Kawaguchi, Saitama 332-0012, Japan} \\
$^d$ {\it RIKEN-FRS, 2-1 Hirosawa, Wako, Saitama 351-0198, Japan} \\
\vskip 3mm
(Received 7 June 2006; revised 19 August 2006)

\end{center}

\begin{abstract}
 Spin torques, that is, 
effects of conduction electrons on magnetization dynamics, 
are calculated microscopically in the first order 
in spatial gradient and time derivative of magnetization. 
 Special attention is paid to the so-called $\beta$-term and the 
Gilbert damping, $\alpha$, in the presence of electrons' 
spin-relaxation processes, which are modeled by quenched magnetic 
impurities. 
 Two types of forces that the electric/spin current exerts 
on magnetization are identified based on a general formula 
relating the force to the torque. 
\end{abstract}


\normalsize 
\vskip 3mm
\noindent
{\large {\bf 1. Introduction}}

 Manipulation of nanoscale magnetization by electric/spin current is of 
recent interest. 
 One well-known effect of the current on magnetization dynamics is the 
spin-transfer torque \cite{Slonczewski,Berger96}. 
 It is expressed as 
$- ({\bm v}_{\rm s} \!\cdot\! {\bm \nabla}) {\bm n}$ 
for a continuous magnetic configuration ${\bm n}$ as described by the 
Landau-Lifshitz-Gilbert (LLG) equation \cite{BJZ98}. 
 Recently, another type of spin torque has been proposed on 
microscopic \cite{WV04,Heide01} and 
phenomenological \cite{Zhang05,Thiaville05,Barnes05} grounds. 
 It has essentially the form 
$-\beta {\bm n} \times ({\bm v}_{\rm s} \!\cdot\! {\bm \nabla}) {\bm n}$, 
and modifies the magnetization dynamics significantly, 
especially in the case of a domain wall by acting as a force 
\cite{Zhang05,Thiaville05,TTKSNF06}. 
 The purpose of the present study is to derive this new torque 
microscopically, by taking the electrons' spin-relaxation processes 
into account, 
and to give a general argument for the force \cite{Berger84,TK04} 
acting on magnetization texture. 
 Some details of the former subject can be found in \cite{KTS}.

\vskip 5mm
\noindent
{\large {\bf 2. Spin Torques}}

 We consider a \lq localized' ferromagnet consisting of 
localized $d$ spins (of magnitude $S$) and conducting $s$ electrons, 
which are coupled each other via the $s$-$d$ exchange interaction, 
$H_{\rm sd} = - M \int d^3x \, {\bm n} \!\cdot\! \hat {\bm \sigma}$. 
 We adopt a continuum description for the localized spin, ${\bm n}$, 
whose dynamics will then be described by the LLG equation 
\begin{eqnarray}
  \dot {\bm n} &=& \gamma_0 {\bm H}_{\rm eff} \times {\bm n} 
            + \alpha_0 \dot {\bm n} \times {\bm n} + {\bm t}_{\rm el}'.
\label{eq:LLG}
\end{eqnarray}
 Here $\gamma_0{\bm H}_{\rm eff}$ and $\alpha_0$ are an effective field 
and a Gilbert damping constant, respectively, coming from the spin part 
of the Hamiltonian. 
 Effects of conduction electrons are contained in the spin torque 
$ {\bm t}_{\rm el} ({\bm r}) 
 =  M {\bm n} ({\bm r}) \times 
    \langle \hat {\bm \sigma} ({\bm r}) \rangle_{\rm n.e.} 
 \equiv  (\hbar S/a^3) \, {\bm t}_{\rm el}' ({\bm r})$, 
which comes from $H_{\rm sd}$. 
 Here $\langle \hat {\bm \sigma} ({\bm r}) \rangle_{\rm n.e.}$ 
is the $s$-electron spin polarization, and $a^3$ is the volume per 
localized $d$ spin. 
 The spin torque is generally expressed as 
\begin{eqnarray}
 {\bm t}_{\rm el}' 
&=& a_0' \dot {\bm n} + ({\bm a}' \!\cdot\! {\bm \nabla})\,  {\bm n} 
  + b_0' \, ({\bm n} \times \dot {\bm n}) 
  + {\bm n} \times ({\bm b}' \cdot\! {\bm \nabla}) \, {\bm n} , 
\label{eq:torque1} 
\end{eqnarray}
in the first order in time derivative and spatial gradients. 
 The coefficients, $a_\mu'$ and $b_\mu'$, can be calculated 
from the linear response of 
$\langle \hat {\bm \sigma} ({\bm r}) \rangle_{\rm n.e.}$ 
to small transverse fluctuations of ${\bm n}$ around the 
uniformly magnetized state 
\cite{TBB}. 
 For a 3D elecron system under the influence of randomly 
distributed nonmagnetic and magnetic impurities 
with bare scattering amplitudes, $u$ and 
$u_{\rm s} {\bm S}_i \!\cdot {\bm \sigma}$, respectively, 
and concentrations, $n_{\rm i}$ and $n_{\rm s}$, 
we obtain 
$a_0' = - \rho_{\rm s} a^3 /2S$, 
${\bm a}' = (a^3 /2eS) {\bm j}_{\rm s}$,  
\begin{eqnarray}
  b_0' 
&=&  - \pi n_{\rm s} u_{\rm s}^2 \!\cdot\! \frac{a^3}{S} \left[\, 
     2 \overline{S_z^2} \nu_\uparrow \nu_\downarrow 
    + \overline{S_\perp^2} (\nu_\uparrow ^2 + \nu_\downarrow^2 ) 
    \right] , 
\label{eq:b0}
\end{eqnarray}
and
\begin{eqnarray}
  {\bm b}' 
 &=&  \frac{\pi n_{\rm s}u_{\rm s}^2}{M} \!\cdot\!  
     \frac{a^3}{2eS} \left[ 
    \bigl( \overline{S_\perp^2} + \overline{S_z^2} \bigr) \nu_+ 
    {\bm j}_{\rm s} 
  + \bigl( \overline{S_\perp^2} - \overline{S_z^2} \bigr) \nu_- 
    {\bm j}_{\rm c}
    \right] .
\label{eq:b}
\end{eqnarray}
 Here 
$\rho_{\rm s} = n_\uparrow - n_\downarrow $ is the equilibrium 
$s$-electron spin polarization, 
${\bm j}_{\rm s} = {\bm j}_\uparrow - {\bm j}_\downarrow 
 \, [ \, \equiv - (2eS/a^3) {\bm v}_{\rm s} \, ]$ 
is the spin current, 
${\bm j}_{\rm c} = {\bm j}_\uparrow + {\bm j}_\downarrow $ 
is the charge current, and 
$\nu_\pm = \nu_\uparrow \pm \nu_\downarrow$ with $\nu_\sigma$ 
being the Fermi-level density of states (DOS) for spin-$\sigma$ 
electrons.
 We have taken a quenched average for the impurity spin direction as 
$ \overline{S_{i,\alpha} S_{j,\beta}} 
= \delta_{ij} \delta_{\alpha\beta} \overline{S_\alpha^2} $ 
with $\overline{S_x^2} = \overline{S_y^2} \equiv \overline{S_\perp^2}$. 
 As seen, only the spin scattering ($u_{\rm s}$), causing spin relaxation, 
contributes to $b_0'$ and ${\bm b}'$, 
and the potential scattering ($u$) does not \cite{com1}. 
 We put $\alpha = - b_0'$ (Gilbert damping) and 
${\bm b}' = - \beta {\bm v}_{\rm s}$ 
(defining $\beta$). 
 In terms of longitudinal 
($\tau_L^{-1} = 4 \pi n_{\rm s} u_{\rm s}^2 \overline{S_\perp^2} 
\nu_+ / \hbar$) 
and transverse 
($\tau_T^{-1} = 2 \pi n_{\rm s} u_{\rm s}^2 
    \bigl( \overline{S_\perp^2} + \overline{S_z^2} \bigr) 
\nu_+ / \hbar$) 
spin-relaxation rates, we have 
\begin{eqnarray}
  \alpha 
 &=&  \frac{a^3 \hbar \nu_+}{4S} \left[ 
    \left( 1 - P_\nu^2 \right) \frac{1}{\tau_T^{\phantom{1}}} 
    + P_\nu^2 \frac{1}{\tau_L^{\phantom{1}}}   \right] , 
\label{eq:alpha2}
\\
  \beta 
 &=&  \frac{\hbar}{2M} \left[ 
    \left( 1 - \frac{P_\nu}{P_j} \right) \frac{1}{\tau_T^{\phantom{1}}} 
    + \frac{P_\nu}{P_j} \frac{1}{\tau_L^{\phantom{1}}}   \right] , 
\label{eq:beta2}
\end{eqnarray}
where $P_j = j_{\rm s} / j_{\rm c} $ is the polarization of 
the current, and $P_\nu = \nu_- / \nu_+ $ is the DOS asymmetry. 
 For \lq\lq isotropic'' impurities with 
$\tau_L^{\phantom{1}} = \tau_T^{\phantom{1}} \equiv \tau_{\rm s}$, 
we have 
$ \alpha = a^3 \hbar \nu_+ /(4S \tau_{\rm s})$, and 
$ \beta = \hbar / (2M\tau_{\rm s})$. 
 Our model was intended to be one of the microscopic realizations of 
the phenomenology of \cite{Zhang05}, 
but now proves to give a different result for $\alpha$, 
namely, with the factor $\nu_+$ replacing $n_0/M$ of \cite{Zhang05}. 
 Furthermore, we could not confirm the relation, $\alpha = \beta$, 
demonstrated in \cite{Barnes05} and \cite{TBB}. 

  For a magnetization varying rapidly in space, 
we have in addition a spatially oscillating torque, 
${\bm t}_{\rm osc}$, due to electron reflection \cite{WV04,TK04}. 
 This torque has the same algebraic form as the $\beta$-term 
but is spatially nonlocal.

 A note on itinerant ferromagnets; in this case, we can also derive 
an LLG equation of the same form, whose coefficients are obtained 
from the above results 
(for \lq localized' ferromagnets) 
by the replacement $2S \to \rho_{\rm s} a^3$ \cite{KTS}. 

\vfill\eject
\vskip 5mm
\noindent
{\large {\bf 3. Forces}}

 The $\beta$-term acts as a force (like magnetic field) for a specific 
case of a rigid domain wall \cite{Zhang05,Thiaville05}. 
 We consider here a generalization of it, and 
derive a general expression for the force acting on 
a fixed but arbitrary  magnetization texture.

 Following a standard procedure in field theory, the energy-momentum 
tensor, especially the momentum density 
$\rho_{{\rm P},i}^{\phantom{\dagger}}$, 
of the magnetization is obtained as 
\begin{eqnarray}
 \rho_{{\rm P},i}^{\phantom{\dagger}} 
 = - \frac{\hbar S}{a^3} (\partial_i \phi ) (\cos\theta -1). 
\end{eqnarray}
 We define the force acting on the magnetization by Newton's 
equation of motion as 
$F_i = (d/dt) \int d^3x \rho_{{\rm P},i}^{\phantom{\dagger}}$. 
 After doing a partial integration (neglecting surface terms), 
and then using the LLG equation, 
$(\hbar S/a^3) \dot {\bm n} = {\bm t}_{\rm tot}$, 
we obtain 
\begin{eqnarray}
 F_i = - \int d^3x \left[ {\bm n} \times (\partial_i{\bm n}) \right] 
  \!\cdot\! {\bm t}_{\rm tot}. 
\label{eq:F}
\end{eqnarray}
 This is a general formula relating the force to the torque. 
 If we write the torque as ${\bm t}_{\rm tot} = {\bm h} \times {\bm n}$ 
in terms of some effective field ${\bm h}$, Eq.(\ref{eq:F}) gives a 
well-known expression, 
$F_i = \int d^3x \, (\partial_i{\bm n}) \cdot {\bm h}$ 
\cite{Thiele73}.

 Each term of the spin torque (\ref{eq:torque1}) exerts a different 
type of force. 
 The spin-transfer torque exerts a \lq transverse' force 
\begin{eqnarray}
 F_{{\rm ST},i} 
 = - \frac{\hbar}{2e} \, j_{{\rm s},\ell} \int d^3x \, 
 \left[ (\partial_i {\bm n}) \times (\partial_\ell {\bm n}) \right] 
  \!\cdot\! {\bm n} , 
\end{eqnarray}
for ${\bm n}$ subtending a finite solid angle (spin chirality). 
 One such example is the magnetic vortex in a ferromagnetic dot, 
whose dynamics driven by electric/spin current is of recent interest 
\cite{SNTKO06,Ishida,Kasai}. 
 The $\beta$-term leads to the force 
\begin{eqnarray}
 F_{{\beta},i} 
 = - \beta \frac{\hbar}{2e} \, j_{{\rm s},\ell} \int d^3x \, 
 (\partial_i {\bm n}) \!\cdot\! (\partial_\ell {\bm n}). 
\end{eqnarray}
 Similar but nonlocal (\lq bilocal') expression results from 
${\bm t}_{\rm osc}$.

 The reactions to the forces, ${\bm F}_{\rm ST}$ and 
${\bm F}_{\beta}$, will affect the $s$-electrons' motion, 
hence their transport, 
and lead to the Hall effect (due to spin chirality) 
\cite{Ye,Ohgushi,Tatara} 
and the dissipative resistivity, respectively. 
 Such a relation between forces and transport coefficients 
has been noted in the study of a domain wall \cite{TK04}, and 
will further be reported elsewhere.

 In summary, 
we have presented a microscopic calculation of the spin torques, 
especially the Gilbert damping and the so-called $\beta$-term, 
on the basis of a microscopic model and controlled approximations. 
 Two types of current-induced forces have been identified based on a 
general relation between the force and the torque.

We would like to thank Y.~Nakatani, H.~Ohno and T.~Ono for 
valuable and informative discussions.

\vfill\eject


\begin{thebibliography}{00}
\bibitem{com1}
 Note that spin-orbit coupling in the bulk band structure is not 
taken into account in the present study.
\end{thebibliography}
\end{document}